\documentclass[12pt,preprint]{aastex}
\def\etal{{\it et~al.~}}
\def\nt{{non-thermal~}}
\def\bsax{{\it BeppoSAX~}}
\def\integral{{\it INTEGRAL~}}
\def\ginga{{\it Ginga~}}

\def\xmm{{\it XMM-Newton~}}

\def\rxte{{\it RXTE~}}

\def\erg{~{\rm erg~ cm}^{-2}\ {\rm s}^{-1}~}

\shorttitle{Confirmation of the HXR excess in Coma}
\shortauthors{R.Fusco-Femiano \etal}

\begin{document}

\newcommand{\lessim}{\ \raise -2.truept\hbox{\rlap{\hbox{$\sim$}}\raise5.truept
    \hbox{$<$}\ }}

\title{Confirmation of non-thermal hard X--ray excess in the Coma
cluster from two epoch observations}

\author{Roberto Fusco-Femiano}
\affil{Istituto di Astrofisica Spaziale e Fisica Cosmica
(IASF/Roma), CNR, via del Fosso del Cavaliere, I--00133 Roma,
Italy} \email{dario@rm.iasf.cnr.it}
\author{Mauro Orlandini}
\affil{IASF/Bologna, C.N.R., via Gobetti 101, I--40129 Bologna,
Italy} \email{orlandini@bo.iasf.cnr.it}
\author{Gianfranco Brunetti, Luigina Feretti}
\affil{Istituto di Radioastronomia , C.N.R., via Gobetti 101,
I--40129 Bologna, Italy} \email{brunetti@ira.cnr.it}
\email{lferetti@ira.cnr.it}
\author{Gabriele Giovannini}
\affil{Dip. di Astronomia, Univ. Bologna, via Ranzani 1, I--40127
Bologna, Italy} \affil{Istituto di Radioastronomia, C.N.R., via
Gobetti 101, I--40129 Bologna, Italy}
\email{ggiovannini@ira.cnr.it}
\author{Paola Grandi}
\affil{IASF/Bologna, C.N.R., via Gobetti 101, I--40129 Bologna,
Italy} \email{grandi@bo.iasf.cnr.it} \and
\author{Giancarlo Setti}
\affil{Dip. di Astronomia, Univ. Bologna, via Ranzani 1, I--40127
Bologna, Italy} \affil{Istituto di Radioastronomia, C.N.R., via
Gobetti 101, I--40129 Bologna, Italy} \email{setti@ira.cnr.it}

\begin{abstract}

We report the hard X--ray spectrum of the \objectname[]{Coma
cluster} obtained using the PDS data of two independent \bsax
observations performed with a time interval of about three years.
In both the spectra a non thermal excess with respect to the
thermal emission is present at a confidence level of $\sim
3.4\sigma$. The combined spectrum obtained by adding up the two
spectra allows a measurement of the excess at the level of $\sim$
4.8$\sigma$ at energies above 20 keV. The analysis of the full
\bsax data set provides a revised \nt X-ray flux which is slightly
lower than that previously estimated (Fusco-Femiano \etal 1999)
and in agreement with that measured by two \rxte observations. The
analysis of the offset fields in our Coma observations provides a
possible flux determination of the BL~Lac object 1ES~1255+244.

\end{abstract}


\section{ Introduction}

In the hierarchical scenario of structure formation, clusters of
galaxies form by the gravitational merger of sub-clusters and
groups. Numerical simulations of large scale structure formation
indicate that this class of objects undergo several merger
processes as they form (West, Villumsen, \& Dekel 1991; Katz \&
White 1993). Cluster mergers are highly energetic events and the
associated large scale shocks and turbulence could provide the
ingredients necessary to the formation of extended radio regions
(radio halos or relics) detected so far in a limited number of
clusters, namely a magnetic field amplification and particle
re-acceleration (Tribble 1993; Roettiger, Burns, \& Stone 1999;
Roettiger, Stone, \& Burns 1999; Cavaliere, Menci, \& Tozzi 1999;
Brunetti \etal 2001; Fujita, Takizawa, \& Sarazin 2003). Clusters
of galaxies with detected diffuse radio emission indeed show
significant evidence of merger activity. The existence of
Mpc-scale radio halos or relics combined with the relatively short
radiative lifetimes of the electrons ($\sim 10^8$ yrs) suggests an
in-situ electron re-acceleration induced by a very recent or
current merger event (Markevitch \& Vikhlinin 2001).

The presence of large radio regions could be related to the origin
of the non-thermal hard X--ray (HXR) emission detected in the Coma
cluster (Fusco-Femiano \etal 1999; Rephaeli, Gruber, \& Blanco
1999; Rephaeli \& Gruber 2002) and \objectname[]{Abell 2256}
(Fusco-Femiano \etal 2000; Rephaeli \& Gruber 2003) by \bsax and
\rxte and, at lower confidence level, by \bsax in
\objectname[]{A754} (Fusco-Femiano \etal 2003). The most likely
interpretation of the non-thermal HXR radiation is inverse Compton
(IC) emission by the same radio synchrotron electrons scattering
the cosmic microwave background (CMB)  photons. However, one
cannot completely exclude the possibility that the non-thermal
emission detected by \bsax and \rxte may be due to the presence of
obscured sources in the field of view of the detectors.  The {\it
BeppoSAX}/MECS images test this possibility only in the cluster
central region of size $\sim 30'$ in radius while the {\it
BeppoSAX}/PDS (Frontera \etal 1997), which is able to detected HXR
radiation in the energy range 15--200 keV, has a larger field of
view (FWHM $\sim 1.3^{\circ}$). The probability to find obscured
sources, like \objectname[]{Circinus} (Matt \etal 1999) very
active at high energies, in the field of view of the PDS is
estimated to be of the order of 10\% (Kaastra 1999; Fusco-Femiano
\etal 2002). Future deep observations by IBIS on-board \integral
with a spatial resolution of $\sim 12'$ can definitely resolve
this uncertainty. A different mechanism given by \nt
bremsstrahlung from supra-thermal electrons formed through the
current acceleration of the thermal gas (En$\ss$lin, Lieu, \&
Biermann 1999; Dogiel 2000; Dolag \& En$\ss$lin 2000; Sarazin \&
Kempner 2000; Blasi 2000; Liang, Dogiel, \& Birkinshaw 2002)
requires an unrealistically high energy input in order to maintain
the HXR emission for more than $10^8$ yrs (Petrosian 2001;2002).
In conclusion, the origin of the detected \nt HXR excesses in some
clusters of galaxies seems to be restricted to a diffuse IC
emission or to the presence of point sources in the field of view
of the PDS.

In this Letter, we present the results of a long \bsax observation
of $\sim$ 300 ksec that confirms the presence of a \nt HXR tail in
the spectrum of the Coma cluster in agreement with the detection
obtained by a previous shorter observation of $\sim$ 91 ksec,
after data re-analysis of the first observation slightly modified
the numerical results reported in Fusco-Femiano \etal 1999, but
not the general conclusions. Finally, we show the combined
spectrum obtained by summing the spectra of the two observations.

Throughout this Letter we assume a Hubble constant of $H_o$ =
50~km~s$^{-1}$~Mpc$^{-1}$~$h_{50}$ and $q_0 = 1/2$, so that an
angular distance of $1^{\prime}$ corresponds to 40.6 kpc
($z_{Coma} = 0.0232$). Quoted confidence intervals are at $90\%$
level, if not otherwise specified.

\section {PDS Data Reduction and Results}

The Coma cluster was observed for the first time in December 1997
for $\sim$91 ksec and re-observed in December 2000 for $\sim$300
ksec. The pointing coordinates of \bsax are at J(2000):
$\alpha:~12^h~ 58^m~ 52^s$; $\delta:~ +27^{\circ}~ 58'~ 54''$. The
total effective exposure times of the PDS in the two observations
were 44.5 ksec and 122.2 ksec, respectively (hereafter OBS1 and
OBS2).

The PDS spectra of both the observations were extracted using the
XAS v2.1 package (Chiappetti \& Dal Fiume 1997). The choice of
using this software is dictated by the non standard pipeline
needed to extract the net count spectra (see below). Because our
Institute (i.e., IASF/Bologna) was in charge of the design,
construction and maintenance of the PDS, and we developed and
tested the XAS package, specifically created to handle the PDS
peculiarities (while the SAXDAS package, used for the standard
analysis, is more suitable for handling imaging instruments, like
MECS and LECS), we felt more confident in using XAS for the PDS
analysis.

Since the source is rather faint in the PDS band ($\sim$5 mCrab in 15--100
keV) a careful check of the background subtraction must be performed. The
background sampling was performed by making use of the default rocking law
of the two PDS collimators that samples ON/+OFF, ON/--OFF fields for each
collimator with a dwell time of 96 sec (Frontera \etal 1997a). When one
collimator is pointing ON source, the other collimator is pointing toward
one of the two OFF positions. Initially, we used the standard procedure to
obtain PDS spectra (Dal~Fiume \etal 1997); this procedure consists of
extracting one accumulated spectrum for each unit for each collimator
position. We then checked the two independently accumulated background
spectra in the two different +/--OFF sky directions, offset by $210'$ with
respect to the on-axis pointing direction (+OFF pointing:
$\alpha:~12^h~58^m~ 57.8^s$; $\delta:~ +24^\circ~ 28'~ 55''.1$ --OFF
pointing: $\alpha:~12^h~58^m~ 47.0^s$; $\delta:~ +31^\circ~ 28'~ 54''.7$).

The comparison between the two accumulated backgrounds (difference
between the +OFF and --OFF count rate spectra) shows that for OBS1
the difference is compatible with zero ($0.044\pm 0.047$ cts/s for
a background level of $21.66\pm 0.02$ cts/s in 15--100 keV), while
for the longer, more sensitive OBS2, there is an excess of
$0.064\pm 0.021$ cts/s (background $16.76\pm 0.01$
cts/s)\footnote{The $\sim$20\% variation in the PDS background is
due to the \bsax orbital decay: the lower orbit for OBS2 increased
the shielding to ambient particles, therefore lowering the diffuse
background.}. A careful check of possible variable sources in the
PDS offset fields lead our attention to the BL~Lac source
\objectname[]{1ES~1255+244}, present in the +OFF field. Luckily,
this same source was observed by \bsax on May 1998 in the
framework of a spectral survey of BL~Lacs by Beckmann \etal
(2002), who, somewhat surprisingly, state "for 1ES~1255+244 there
are no PDS data". Indeed, we retrieved the raw data from the ASI
Scientific Data Center, extracted the PDS spectrum (the background
has to be evaluated only on one offset field because the other is
pointed exactly on Coma --- the two sources are contaminating each
other!) and found that the source is quite faint, consistent with
zero flux being detected. Because of the very short exposure time
($\sim$ 3~ksec) it is only possible to give a 2$\sigma$ upper
limit of 0.26 cts/s in 15--100 keV, corresponding to 1.6 mCrab,
however compatible with the background excess measured in OBS2.

It is worth noticing that it is possible to derive a more
stringent upper limit on the X--ray flux from 1ES~1255+44 at the
time of our Coma OBS2 by assuming that all the contamination in
the +OFF field comes from this source. It is straightforward to
show that the $0.064\pm 0.021$ cts/s excess translates into
$428\pm 424$ net counts when one takes into account the much
shorter BL~Lac observation (3340 sec) compared to our OBS2 and an
upward factor two correction due to the $\sim 40'$ off-axis
position of the source, corresponding to a 50\% intensity
reduction because of the triangular collimator response (Frontera
\etal 1997b). At its face value this is consistent with the run of
PDS counts as compared to the MECS counts in the BL~Lac sample
studied by Beckmann \etal (2002), barring time variability
effects.

Returning to the Coma observations, to remain on the safe side we
decided to exclude the +OFF field in the background evaluation,
and consider only the --OFF field as the "un-contaminated"
background for both the Coma observations. Moreover, just in the
center of the +OFF field is also present the extremely weak ROSAT
source \objectname[]{RX~J125847.1+242741}. However, in section 3
we also report the level of confidence of the \nt excess
considering the average of the measured backgrounds in the two
positions.

The observed count rate of OBS1 is 0.78$\pm$0.03 cts/s in the
15--100 keV energy range, at a confidence level of $\sim
26\sigma$. In the first analysis the \nt excess with respect to
the thermal bremsstrahlung emission reported by the PDS was at the
confidence level of $\sim 4.5\sigma$. The derived \nt flux was
$\sim 2.2\times 10^{-11}\erg$ in the 20--80 keV energy band
(assuming a photon index $\Gamma_X$ = 1.5). By relating the radio
and the \nt fluxes it was then possible to estimate a
volume-averaged intracluster magnetic field $B_X\sim 0.15\mu G$,
using only observables (Fusco-Femiano \etal 1999). It should be
pointed out that a re-analysis of these data has evidenced a
trivial mistake in the previous data analysis (summing three
spectra, one of them was summed twice), so that the correct
spectrum obtained from OBS1 shows an excess with respect to the
thermal component with the average gas temperature measured by
\ginga (8.11$\pm$0.07, 90\%; David \etal 1993) at a somewhat lower
confidence level of $\sim 3.4\sigma$ (see Table~1). The fit with a
single temperature gives $\sim 9.9^{+1.3}_{-1.1}$ keV, above the
average gas temperature measured by {\it Ginga} (with a field of
view comparable to that of the PDS), implying the presence of a
second spectral component. The fit with two thermal components
(one fixed at 8.1 keV) requires an unrealistic second temperature
($>$ 50 keV) that strongly supports a \nt mechanism for the
additional component present in the spectrum of the Coma cluster.
If we consider a power-law for the second component, the PDS data
are not able to fix the photon index, but the \nt flux is rather
stable against index variations. We assume a photon index
$\Gamma_X$ = 2.0 to derive the \nt flux that results to be
$(2.3\pm 1.0)\times 10^{-11}\erg$ in the 20--80 keV energy range.

The observed count rate of OBS2 is 0.72$\pm$0.02 cts/s in the
15--100 keV energy range, at the confidence level of $\sim
36\sigma$. At energies above 20 keV the spectrum shows an excess
with respect to the thermal emission (kT = 8.1 keV) at a
confidence level of $\sim 3.4\sigma$ (see Table~1). The fit with a
single temperature gives $\sim 9.5^{+0.8}_{-0.6}$ keV. Also OBS2
indicates the presence of an additional spectral feature and also
in this case the fit with a second thermal component requires
unrealistic values for the temperature. The \nt flux
($\Gamma_X$=2.0) is $(1.3^{+0.5}_{-0.6})\times 10^{-11}\erg$ in
the band 20--80 keV, consistent with the flux reported in OBS1.
The \nt fluxes are (marginally) consistent also at a 68\%
confidence level:  $(2.3\pm 0.7)\times 10^{-11}\erg$ in the first
observation and $(1.3^{+0.3}_{-0.4})\times 10^{-11}\erg$ in the
second one.

The combined spectrum is obtained by summing the spectra of the
two observations (see Fig.~1). The total count rate is $0.740\pm
0.017$ cts/s in the 15--100 keV energy range, at the confidence
level of $\sim 44\sigma$. At energies $>$20 keV the HXR excess is
at the confidence level of $\sim 4.8\sigma$ (see Table~1). Even
the inclusion of a 1\% systematic to the data, necessary for
sources with high S/N ratio but not for faint sources like Coma
(Frontera \etal 1997b), does not change the significance of our
\nt HXR excess. The fit with a single thermal component gives
9.7$\pm0.6$ keV, well above the average gas temperature measured
by {\it Ginga}, with a statistically unacceptable $\chi^2$ value
(=2.1 for 8 d.o.f.). The presence of a second component is more
evident from the $\chi^2$ value that has a significant decrement
when a second component, a power law, is added to the thermal
component with kT=8.1 keV. The improvement passing from the first
model ($\chi^2_\nu$ = 4.10 for 9 d.o.f.) to the second one
($\chi^2_\nu$ = 1.2 for 7 d.o.f.) is significant at more than
99.4\% confidence level, according to the F-test. Also, the
combined spectrum cannot be fitted with a second thermal component
unless an unrealistic value for the temperature is assumed, thus
supporting the \nt origin for this additional spectral feature.
The \nt flux for $\Gamma_X$=2.0 is $(1.5\pm0.5)\times
10^{-11}\erg$ in the 20--80 keV energy range and it is rather
stable assuming reasonable different values of the photon index.
In fact, for $\Gamma$ =1.5 the flux is $\sim$ 6\% lower and for
$\Gamma$ = 2.5 is $\sim$ 15\% higher.

\section{Discussion}

Non-thermal HXR emission has been reported in two \bsax
observations of the Coma cluster performed with a time interval of
about three years. Both the observations indicate the presence of
a \nt excess with respect to the thermal emission at a confidence
level of $\sim 3.4\sigma$. The combined spectrum obtained by
adding up the two spectra gives an excess at the confidence level
of $\sim 4.8\sigma$. The spectra of the two observations have been
obtained using only the uncontaminated background-accumulated
pointing at the [--OFF] field. However, by considering the average
of the background measurements in the two sky directions in both
the observations, the excess is still significant at the level of
$\sim 3.9\sigma$ (observed count rate = 0.324$\pm$0.013 cts/s,
model predicted rate = 0.273 cts/s). The \nt fluxes measured in
the two observations are consistent at the 90\% confidence level
and marginally at the 68\% c.l..

The \nt flux derived from the combined spectrum,
$(1.5\pm0.5)\times 10^{-11}\erg$ in the 20--80 keV energy range,
is consistent with our previously published detection of $(2.2\pm
0.8)\times 10^{-11}\erg$ and with the value of $(1.2\pm0.3)\times
10^{-11}\erg$ measured by \rxte in the same energy band (Rephaeli,
Gruber, \& Blanco 1999) and confirmed by a second deeper
observation (Rephaeli \& Gruber 2002). The \nt flux value and the
relative c.l. of the excess have been obtained using the \ginga
measurement of 8.11$\pm$0.07 keV (David \etal 1993) for the
average gas temperature that is in good agreement with the \xmm
determination (Arnaud \etal 2001) of 8.25$\pm$0.10 keV in the
central region (R$< 10'$) of the cluster. In fact, all the X-ray
observations of Coma (Hughes, Fabricant, \& Gorenstein 1998a;
Hughes \etal 1988b; Watt \etal 1992; Hughes \etal 1993) have
indicated a clear pattern: the larger the field of view, the lower
the measured temperature. Besides, \rxte reports a best-fit
temperature of 7.90$\pm$0.03 keV (Rephaeli \& Gruber 2002) in a
field of view of $\sim 1^{\circ}$ comparable to that of the PDS.
However, also considering kT=8.25 keV for the average gas
temperature in the field of view of the PDS, the \nt excess is at
the level of $\sim 4.6\sigma$ (observed count rate =
0.349$\pm$0.015 cts/s, model predicted rate = 0.280 cts/s) and the
derived \nt flux has a negligible variation.

Recent PDS data analysis of Coma performed with the SAXDAS
software have lead to controversial results: an analysis of both
observations has not reported evidence for a \nt excess (Rossetti
\& Molendi 2003), while the analysis of the first one (OBS1) by
Nevalainen \etal (2003) confirms our published HXR detection,
albeit at lower confidence level for the systematic uncertainties
of their work. A systematic comparison between PDS spectra
extracted by means of the two software packages (XAS, used here,
and SAXDAS) is under way (Landi \etal, 2003). Preliminary results
on the analysis performed on sources of different luminosities
show that the spectral parameters do not change when computed with
different packages. On the other hand, the errors associated to
the spectral parameters are smaller when using XAS. This effect is
more accentuated for faint sources. We suspect that this could be
due to differences in filtering of good data and/or in the
spectral equalization (i.e., conversion from spectral channels to
energy channels) that for SAXDAS is performed {\em after} summing
the four PDS units while for XAS is performed {\em before}. These
effects will be discussed into details in a forthcoming paper (Landi \etal in
preparation).

As discussed in the Introduction, the likely origin of the \nt HXR
excesses detected by \bsax and \rxte is IC emission by the same
relativistic electrons responsible for the diffuse radio emission
scattering the CMB photons, as predicted in the 1970`s (see Perola
\& Reinhardt 1972; Rephaeli 1979). The probability to find
obscured sources in the field of view of the PDS may be lower than
$\sim$ 10\%, as estimated by Kaastra \etal (1999) and
Fusco-Femiano \etal (2002), considering that \nt HXR emission has
been detected in at least two clusters, Coma and A2256, both
showing extended radio emission. A detailed search by Nevalainen
\etal (2003) excludes a significant contamination from obscured
AGN present in the FOV of Coma and A2256 and supports an
indication for an extended non thermal emission. If we take into
account the estimated contribution by AGN in Coma:
(9$\pm$6)$\times$ 10$^{-3}$ cts/s in the 20--80 keV energy range,
the confidence level of the \nt excess is in the interval $\sim
(3.9-4.7)\sigma$.

In the framework of the IC model the combination of the radio and
\nt X--ray fluxes allows an estimate for a volume-averaged
intracluster magnetic field $B_X$ of $\sim 0.2\mu G$ (see
Fusco-Femiano \etal 1999). This value seems to be in contrast with
the line of sight magnetic field derived from the Faraday rotation
of polarized radiation of sources through the ICM ($B_{FR} \sim
6\mu G$; Feretti \etal 1995). Newman, Newman \& Rephaeli (2002)
have recently pointed out that many and large uncertainties are
associated with the determination of $B_{FR}$ (see also Govoni
\etal 2002 and Govoni \& Murgia 2003). However, this discrepancy
can be attenuated by considering models that include the effects
of more realistic electron spectra, spatial profiles of the
magnetic fields and anisotropies in the pitch angle distribution
of the electrons (Goldshmidt \& Rephaeli 1993; Brunetti \etal
2001; Petrosian 2001; Kuo, Hwang, \& Ip 2003). The present value
of the \nt HXR flux is slightly lower than that reported in
Fusco-Femiano \etal (1999) favoring a central magnetic field
strength of 1--2 $\mu$G in the {\it two-phase} model of Brunetti
\etal (2001), more consistent with the $B_{FR}$ values.

The alternative between primary and secondary electrons as
responsible for non-thermal phenomena in clusters of galaxies is
discussed in many papers. Primary electrons may be injected in the
ICM of the Coma cluster by some processes (starbursts, AGNs,
shocks, turbulence) during a first phase and re-accelerated during
a second phase (Brunetti \etal 2001). Secondary electrons may be
due to decay of charged pions generated in cosmic ray collisions
within the ICM (Dennison 1980; Blasi \& Colafrancesco 1999; Dolag
\& En$\ss$lin 2000; Miniati \etal 2001; Miniati 2003). Radio and
HXR spectral properties of Coma provide observational constraints
able to discriminate between these two different populations of
electrons (Brunetti 2002). In particular, the derived
volume-averaged intracluster magnetic field of $\sim 0.2\mu G$
implies relativistic electrons at energies $\gamma\sim 10^4$ to
explain the observed diffuse synchrotron emission. At these
energies IC losses may determine a cutoff in the spectrum of the
accelerated electrons as supported by the radio spectral cutoff
observed in Coma (Deiss \etal 1997). The cutoff in the electron
spectrum may be naturally accounted for in the context of
re-acceleration models, while it is not expected if the radio
emission is due to a continuous production of secondary electrons.
More recently, a radio spectral cutoff has been found also in the
case of A754 by relating the VLA observation at 1.4 GHz (Bacchi
\etal 2003) to the observations of Kassim \etal (2001) at lower
frequencies. This cluster also shows \nt HXR radiation detected at
a confidence level slightly above 3$\sigma$ by \bsax
(Fusco-Femiano \etal 2003) and the derived value of the magnetic
field is of the same order of that determined in Coma. The PDS
detection should be confirmed by a deeper observation with imaging
instruments for the presence of the radio galaxy 26W20 located at
a distance of $\sim 27'$ from the \bsax pointing. IBIS on-board
\integral with its spatial resolution of $\sim 12'$ has the
possibility to eliminate this ambiguity and to detect the excess
at a higher confidence level with respect to that obtained by {\it
BeppoSAX}.

\section{Acknowledgments}

We wish to thank F.~Frontera for stimulating discussions and the
referee for the useful suggestions. This research has made use of
the SIMBAD database, operated at CDS, Strasbourg, France, and of
data retrieved from the ASI Scientific Data Center operated at the
ESA establishment of ESRIN, Frascati, Italy.

\clearpage

\begin{deluxetable}{ccccccc}
\tablecolumns{7}
\tablewidth{0pc}
\tablecaption{Non-thermal HXR excess in 20--80 keV PDS observations\tablenotemark{a}\label{tab1}}
\tablehead{%
\colhead{} & \colhead{Epoch} & \colhead{PDS exposure} & \colhead{Observed rate} & \colhead{Predicted rate} & \colhead{Excess} &
  \colhead{Flux\tablenotemark{b}} \\
\colhead{} & \colhead{}      & \colhead{(ksec)}       & \colhead{(cts/s)}       & \colhead{(cts/s)}        & \colhead{(c.l.)} &
  \colhead{}
}

\startdata
OBS1 & Dec 1997 &  44.5 & $0.390\pm 0.033$ & 0.278 & 3.4$\sigma$ & $2.3\pm 1.0$ \\
OBS2 & Dec 2000 & 122.2 & $0.333\pm 0.017$ & 0.275 & 3.4$\sigma$ & $1.3_{-0.6}^{+0.5}$ \\
     & Combined    & 166.7 & $0.349\pm 0.015$ & 0.276 & 4.8$\sigma$ & $1.5\pm 0.5$ \\
\enddata

\tablecomments{Quoted errors at 90\% confidence level for a single
parameter.} \tablenotetext{a}{\ Excess with respect to a 8.1 keV
(David \etal 1993) thermal bremsstrahlung component for energies
above 20 keV (see text for details).} \tablenotetext{b}{\ In units
of $10^{-11} \erg$. A photon index of 2 was used to derive the
flux (see text for details).}

\end{deluxetable}

\clearpage

\begin{figure}
\rotatebox{-90}{
\epsscale{0.7}
\plotone{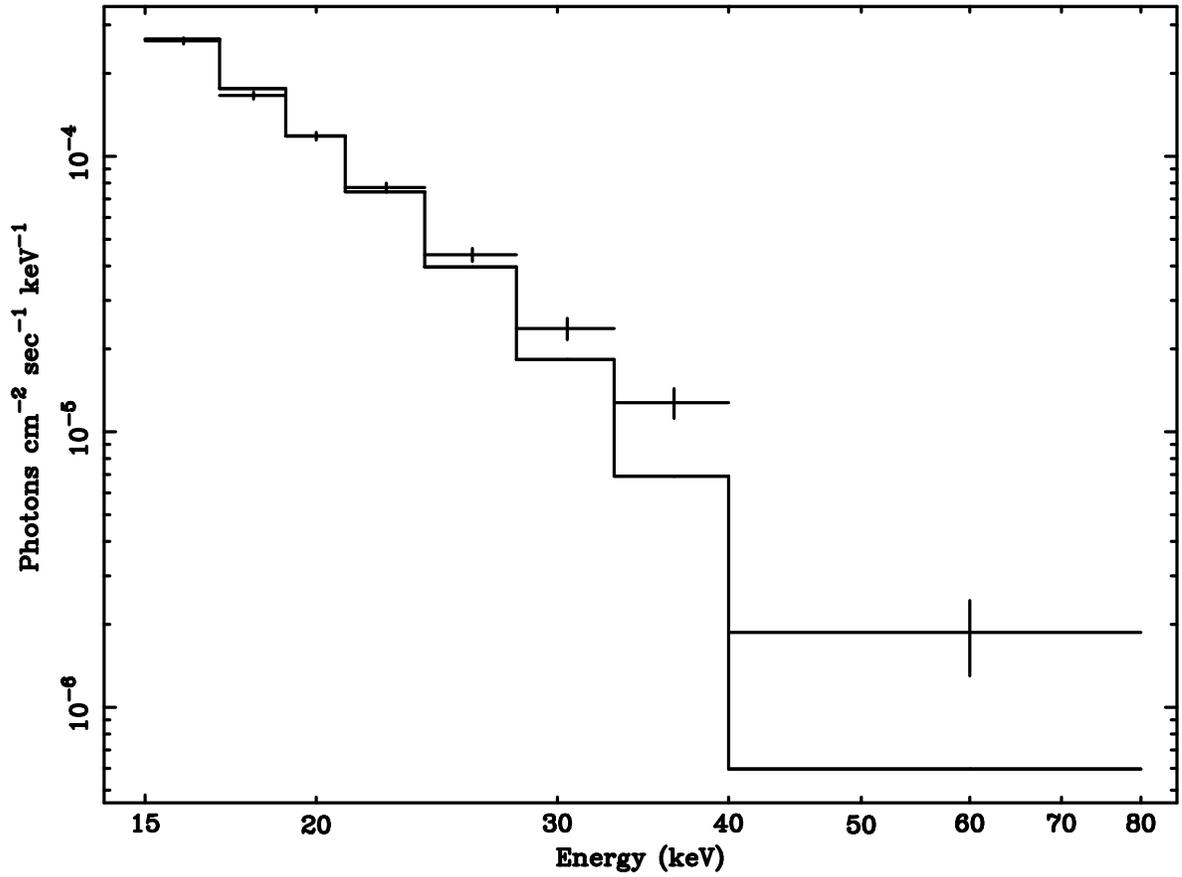}}
\vspace{2cm}
\caption{Coma cluster
--- PDS data. The continuous line represents the best fit with a
thermal component at the average cluster gas temperature of 8.1
keV (David \etal 1993). The errors bars are quoted at the
1$\sigma$ level.\label{fig1}}
\end{figure}

\end{document}